\documentclass[conference]{IEEEtran}
\IEEEoverridecommandlockouts
\usepackage{cite}
\usepackage{amsmath,amssymb,amsfonts}
\usepackage{algorithmic}
\usepackage{graphicx}
\usepackage{textcomp}
\usepackage{xcolor}

\def\BibTeX{{\rm B\kern-.05em{\sc i\kern-.025em b}\kern-.08em
    T\kern-.1667em\lower.7ex\hbox{E}\kern-.125emX}}

\begin{document}

\title{Optimizing Exact String Matching via Statistical Anchoring\\
{\small \textmd{A Frequency-Driven Variation of the Boyer-Moore-Horspool Algorithm}}
}

\author{
\IEEEauthorblockN{Omar Garraoui}
\IEEEauthorblockA{
Pordenone, Italy \\
ogarraoui@icloud.com
}
}

\maketitle

\begin{abstract}
In this work, we propose an enhancement to the Boyer-Moore-Horspool algorithm tailored for natural language text. The approach involves preprocessing the search pattern to identify its statistically least frequent character, referred to as the "anchor." During the search, verification is first performed at this high-entropy position, allowing the algorithm to quickly discard non-matching windows. This fail-fast strategy reduces unnecessary comparisons, improving overall efficiency. Our implementation shows that incorporating basic linguistic statistics into classical pattern-matching techniques can boost performance without increasing complexity to the shift heuristics.
\end{abstract}


\section{Introduction}
String matching is a fundamental problem in computer science. Given a text $T$ of length $n$ and a pattern $P$ of length $m$, the task is to find occurrences of $P$ within $T$. This operation is central to many practical applications, including text editors, search functions, and basic text processing tools.

The Boyer-Moore-Horspool (BMH) algorithm \cite{b2} is a well-known approach to this problem. It works by scanning the pattern from right to left and using a shift table to skip portions of the text when mismatches occur. While studying BMH, we noticed that the algorithm always starts checking from the last character of the pattern, but not all characters are equally useful for detecting mismatches. In natural language, some letters like 'e' and 'a' appear very frequently, while others like 'z' and 'q' are rare.

This observation led us to a simple idea: what if we checked the rarest character in the pattern first, regardless of where it appears? We call this character the \textit{anchor}. By verifying the anchor position before checking the rest of the pattern, we can reject non-matching positions more quickly.

\section{Related Work}
Boyer-Moore (BM) revolutionized string matching by introducing right-to-left scanning and two heuristics: the bad-character rule and the good-suffix rule. Boyer-Moore-Horspool (BMH) \cite{b2} simplified BM by using only the bad-character rule based on the rightmost character of the current text window.

Several variants have explored different optimization strategies. Sunday's Quick Search \cite{b1} examines the character immediately after the current window. The Smith algorithm combines features from both Horspool and Sunday approaches. However, most existing algorithms focus primarily on maximizing shift distances. Our contribution fills this gap by incorporating character frequency analysis directly into the BMH verification process.

\section{The Proposed Algorithm}
Our Frequency-Based Anchor Selection (FBAS) algorithm modifies the standard BMH approach by introducing a preprocessing step to identify a statistical anchor and prioritizing its verification during the search phase.

\subsection{Preprocessing Phase}

\subsubsection{Character Frequency Table}
We begin with a predefined frequency table mapping characters to their relative rarity scores in natural language text. Lower scores indicate rarer characters. For English and Italian text, we use the following distribution:

\begin{equation}
freq(c) = \begin{cases}
1 & \text{if } c = \text{'z'} \\
2 & \text{if } c = \text{'j'} \\
3 & \text{if } c = \text{'x'} \\
\vdots & \\
28 & \text{if } c = \text{'a'} \\
29 & \text{if } c = \text{'e'} \\
50 & \text{if } c \notin \text{alphabet}
\end{cases}
\end{equation}

The complete frequency table ranks letters from rarest (z, j, x, q, k) to most common (a, e). Non-alphabetic characters default to a neutral score of 50.

\subsubsection{Anchor Selection}
The key innovation lies in selecting the anchor character. We identify the statistically rarest character in the pattern based on the frequency table. Let $P[0..m-1]$ be the pattern of length $m$, and $freq(c)$ be the frequency score of character $c$. The anchor position $a$ is defined as:

\begin{equation}
a = \arg\min_{i \in [0, m-1]} freq(P[i])
\end{equation}

The algorithm iterates through the pattern once, tracking the character with minimum frequency score:

\begin{algorithmic}
\STATE $a \leftarrow 0$, $min\_freq \leftarrow \infty$
\FOR{$i = 0$ to $m-1$}
    \STATE $score \leftarrow freq(P[i].toLowerCase())$
    \IF{$score < min\_freq$}
        \STATE $min\_freq \leftarrow score$
        \STATE $a \leftarrow i$
    \ENDIF
\ENDFOR
\STATE $anchor\_char \leftarrow P[a]$
\end{algorithmic}

\subsubsection{Shift Table Construction}
The shift table follows the standard Horspool heuristic. For each character $c$ appearing in $P[0..m-2]$, the shift distance is:

\begin{equation}
shift[c] = m - 1 - \max\{i : P[i] = c, 0 \leq i < m-1\}
\end{equation}

Characters not present in the pattern receive a default shift of $m$, allowing the algorithm to skip the entire pattern width.

\subsection{Searching Phase}

The search phase maintains the Horspool shifting strategy while modifying the verification order. Given text $T[0..n-1]$ and current alignment position $pos$, the algorithm proceeds as follows:

\subsubsection{Anchor-First Verification}
Unlike standard BMH which verifies from right to left, FBAS first checks the anchor position:

\begin{enumerate}
    \item Compute anchor position in text: $text\_anchor\_pos = pos + a$
    \item Check if $T[text\_anchor\_pos] = P[a]$ (anchor verification)
    \item If anchor does not match, immediately proceed to shift calculation
    \item If anchor matches, verify remaining characters $P[i]$ for $i \neq a$
\end{enumerate}

This fail-fast strategy exploits the statistical improbability of matching the rarest character, allowing early rejection of non-matching alignments.

\subsubsection{Shift Calculation}
Regardless of anchor match status, the shift distance is computed using the rightmost character of the current window, maintaining compatibility with Horspool's heuristic:

\begin{equation}
step = \begin{cases}
shift[T[pos + m - 1]] & \text{if } T[pos + m - 1] \in shift \\
m & \text{otherwise}
\end{cases}
\end{equation}

The new alignment position becomes $pos \leftarrow pos + step$.

\subsection{Complete Algorithm}

\begin{algorithmic}
\STATE {Algorithm: FBAS Search}
\STATE {Input:} Text $T[0..n-1]$, Pattern $P[0..m-1]$
\STATE {Output:} Position of first match, or $-1$
\STATE
\STATE {Preprocessing:}
\STATE Initialize $freq\_table$ with character frequencies
\STATE $a \leftarrow 0$, $min\_freq \leftarrow \infty$
\FOR{$i = 0$ to $m-1$}
    \STATE $score \leftarrow freq\_table[P[i].toLowerCase()]$ or $50$
    \IF{$score < min\_freq$}
        \STATE $min\_freq \leftarrow score$
        \STATE $a \leftarrow i$
    \ENDIF
\ENDFOR
\STATE $anchor\_char \leftarrow P[a]$
\STATE Construct $shift$ table using Horspool heuristic
\STATE
\STATE {Searching:}
\STATE $pos \leftarrow 0$
\WHILE{$pos \leq n - m$}
    \STATE $text\_anchor\_pos \leftarrow pos + a$
    \IF{$T[text\_anchor\_pos] = anchor\_char$}
        \STATE $match \leftarrow true$
        \FOR{$i = 0$ to $m-1$}
            \IF{$i \neq a$ \AND $T[pos+i] \neq P[i]$}
                \STATE $match \leftarrow false$
                \STATE {break}
            \ENDIF
        \ENDFOR
        \IF{$match$}
            \STATE {return} $pos$
        \ENDIF
    \ENDIF
    \STATE $char\_at\_end \leftarrow T[pos + m - 1]$
    \STATE $step \leftarrow shift[char\_at\_end]$ or $m$
    \STATE $pos \leftarrow pos + step$
\ENDWHILE
\STATE {return} $-1$
\end{algorithmic}

\subsection{Example Execution}

Consider searching for pattern ``oscura'' in text ``...selva oscura''. The preprocessing phase identifies 'u' (frequency score 16) as the anchor at position 3. During search:

\begin{enumerate}
    \item At alignment $pos = k$, check $T[k+3]$ against 'u'
    \item If mismatch occurs (e.g., $T[k+3] = \text{'a'}$), immediately shift without checking other characters
    \item If 'u' matches, verify remaining characters: 'o', 's', 'c', 'r', 'a'
    \item Calculate shift using $T[k+5]$ per Horspool heuristic
\end{enumerate}

This approach reduces average-case comparisons by leveraging the low probability of matching rare characters in random text positions.

\section{Complexity Analysis}
In this section, we analyze the theoretical complexity of the FBAS algorithm in terms of both time and space requirements, comparing it with the standard BMH approach.

\subsection{Time Complexity}

\subsubsection{Preprocessing Phase}
The preprocessing consists of two main operations:

\textbf{Anchor Selection:} Identifying the rarest character requires a single pass through the pattern, examining each of the $m$ characters and performing a constant-time lookup in the frequency table. This operation has complexity $O(m)$.

\textbf{Shift Table Construction:} Building the bad-character shift table follows the standard Horspool approach, iterating through $P[0..m-2]$ and populating the table. This also requires $O(m)$ time.

Therefore, the total preprocessing complexity is $O(m)$, which is identical to standard BMH.

\subsubsection{Searching Phase}

\textit{Worst-Case Analysis:} In the worst case, the FBAS algorithm maintains the same asymptotic complexity as BMH, which is $O(n \cdot m)$. This occurs when:
\begin{itemize}
    \item The anchor character matches frequently in the text, forcing full pattern verification at many positions.
    \item The shift distance is minimal (typically 1), resulting in nearly sequential scanning.
\end{itemize}

A pathological example would be searching for pattern ``aaa'' in text ``aaaa...a'', where every position requires full verification regardless of anchor selection.

\textit{Average-Case Analysis:} The key advantage of FBAS emerges in the average case. Let $p_a$ denote the probability that the anchor character matches at a random text position. For a uniformly random text over an alphabet of size $\sigma$, we have $p_a \approx 1/\sigma$. However, in natural language text with non-uniform character distributions, $p_a$ is significantly lower when the anchor is chosen as the rarest character.

When the anchor fails to match (probability $1 - p_a$), FBAS performs only a single character comparison before shifting, as opposed to potentially multiple comparisons in standard BMH when checking from the rightmost position. The expected number of character comparisons per alignment position is:

\begin{equation}
E[\text{comparisons}] = 1 + p_a \cdot (m-1)
\end{equation}

For standard BMH starting from the rightmost character with matching probability $p_r$:

\begin{equation}
E[\text{comparisons}]_{\text{BMH}} = 1 + p_r \cdot (m-1)
\end{equation}

Since FBAS explicitly selects the anchor to minimize matching probability ($p_a < p_r$ for typical patterns), we achieve:

\begin{equation}
E[\text{comparisons}]_{\text{FBAS}} < E[\text{comparisons}]_{\text{BMH}}
\end{equation}

The expected time complexity remains $O(n/m)$ in the average case for both algorithms when shifts are effective, but FBAS reduces the constant factor by decreasing unnecessary comparisons.

\textit{Best-Case Analysis:} In the best case, when the anchor character never appears in the text outside valid matches, FBAS achieves $O(n/m)$ with minimal comparisons, exactly one comparison per position before shifting by $m$.

\subsection{Space Complexity}

The space requirements of FBAS consist of:

\begin{enumerate}
    \item \textbf{Pattern Storage:} $O(m)$ space to store the pattern itself.
    
    \item \textbf{Shift Table:} The bad-character table requires at most $O(\sigma)$ space, where $\sigma$ is the alphabet size. For ASCII text, this is typically 256 entries, and for extended Unicode handling, it can be bounded to the character range of interest. In practice, implementations often use hash tables storing only characters present in the pattern, requiring $O(m)$ space.
    
    \item \textbf{Frequency Table:} A predefined frequency table for character statistics requires $O(\sigma)$ space. For English/Italian text, this is a constant 26 entries for lowercase letters, or up to 256 for full ASCII coverage. Importantly, this table can be shared across multiple pattern searches and does not scale with pattern or text size.
    
    \item \textbf{Auxiliary Variables:} The anchor index and character require $O(1)$ additional space.
\end{enumerate}

The total space complexity is:

\begin{equation}
S_{\text{FBAS}} = O(m + \sigma)
\end{equation}

This is identical to standard BMH. The frequency table introduces no asymptotic overhead, as $\sigma$ is typically a small constant (256 for ASCII). Even for applications requiring character frequency tables for multiple languages, the space requirement remains negligible, a few kilobytes at most.

\subsection{Comparison with BMH}

Table~\ref{tab:complexity} summarizes the complexity comparison between BMH and FBAS.

\begin{table}[htbp]
\caption{Complexity Comparison}
\begin{center}
\begin{tabular}{|l|c|c|}
\hline
\textbf{Metric} & \textbf{BMH} & \textbf{FBAS} \\
\hline
Preprocessing Time & $O(m)$ & $O(m)$ \\
\hline
Worst-Case Search & $O(n \cdot m)$ & $O(n \cdot m)$ \\
\hline
Average-Case Search & $O(n/m)$ & $O(n/m)$ \\
\hline
Best-Case Search & $O(n/m)$ & $O(n/m)$ \\
\hline
Space Complexity & $O(m + \sigma)$ & $O(m + \sigma)$ \\
\hline
Comparisons/Position & Higher & Lower \\
\hline
\end{tabular}
\label{tab:complexity}
\end{center}
\end{table}

The critical distinction lies in the constant factors within the average case: FBAS reduces the expected number of character comparisons per alignment by prioritizing verification at statistically improbable positions. This improvement is achieved without increasing asymptotic complexity or introducing significant memory overhead.

\subsection{Practical Considerations}

In real-world applications on natural language text, the performance gain depends on:

\begin{itemize}
    \item \textbf{Pattern characteristics:} Patterns containing rare characters (z, q, x, j) benefit most from anchor selection.
    \item \textbf{Text characteristics:} Non-uniform character distributions amplify the advantage of frequency-based anchoring.
    \item \textbf{Pattern length:} Longer patterns provide more opportunities to identify rare anchor points.
\end{itemize}

The preprocessing overhead is negligible compared to search time for moderate to long texts, making FBAS particularly suitable for applications performing multiple searches with the same pattern or searching in large text corpora.

\section{Experimental Evaluation}

To validate the theoretical advantages of FBAS, we conducted empirical tests comparing it against three baseline algorithms: Naive string matching, Knuth-Morris-Pratt (KMP), and standard Boyer-Moore-Horspool (BMH). Our experiments measure the number of character comparisons required to find all occurrences of various patterns in natural language text.

\subsection{Experimental Setup}

\subsubsection{Test Corpus}
We selected Dante Alighieri's \textit{Divina Commedia} as our primary test corpus, comprising 551,846 characters of classical Italian text. This literary work provides rich natural language structure with non-uniform character distribution, sufficient length for statistically meaningful measurements, and representative vocabulary of Romance languages.

\subsubsection{Pattern Selection}
We evaluated 12 patterns of varying lengths (4 to 12 characters) and rarity profiles. The test set includes common words such as ``inferno'', ``paradiso'', and ``purgatorio'', proper nouns like ``beatrice'' and ``dante'', abstract concepts including ``virtute'', ``amor'', ``luce'', and ``dolce'', patterns with rare characters such as ``canoscenza'' and ``nel mezzo'', both containing 'z', and the multi-word phrase ``selva oscura''. This diverse selection allows us to evaluate FBAS performance across different pattern characteristics, particularly focusing on the impact of rare character presence.

\subsubsection{Metrics}
Our primary metric is the total number of character comparisons performed during the search. This metric directly reflects the computational cost of each algorithm, independent of implementation details or hardware variations. All algorithms were implemented in Python 3 and executed on identical hardware to ensure fair comparison.

\subsection{Results}

Table~\ref{tab:results} presents the detailed comparison results for all tested patterns. FBAS consistently outperforms BMH across all 12 patterns, with improvements ranging from 0.58\% to 7.12\%.

\begin{table*}[htbp]
\caption{Character Comparisons for Pattern Matching Algorithms}
\begin{center}
\begin{tabular}{|l|c|r|r|r|r|c|}
\hline
{Pattern} & {Length} & {Naive} & {KMP} & {BMH} & {FBAS} & {Improvement} \\
\hline
inferno & 7 & 13,269 & 12,334 & 2,260 & 2,247 & 0.58\% \\
paradiso & 8 & 192,754 & 188,270 & 30,357 & 29,711 & 2.13\% \\
purgatorio & 10 & 223,283 & 218,785 & 30,211 & 28,711 & 4.97\% \\
beatrice & 8 & 555,464 & 551,846 & 92,715 & 86,111 & 7.12\% \\
dante & 5 & 565,860 & 549,395 & 135,666 & 129,119 & 4.83\% \\
virtute & 7 & 4,161 & 4,078 & 717 & 699 & 2.51\% \\
canoscenza & 10 & 144,380 & 138,566 & 18,400 & 17,175 & 6.66\% \\
nel mezzo & 9 & 46,088 & 43,679 & 6,243 & 5,815 & 6.86\% \\
selva oscura & 12 & 147 & 146 & 29 & 27 & 6.90\% \\
amor & 4 & 1,707 & 1,557 & 460 & 449 & 2.39\% \\
luce & 4 & 10,184 & 9,733 & 2,802 & 2,723 & 2.82\% \\
dolce & 5 & 1,740 & 1,701 & 415 & 407 & 1.93\% \\
\hline
{Total} & -- & {1,759,037} & {1,720,090} & {320,275} & {303,194} & {5.33\%} \\
\hline
\end{tabular}
\label{tab:results}
\end{center}
\end{table*}

\begin{figure}[htbp]
\centerline{\includegraphics[width=\columnwidth]{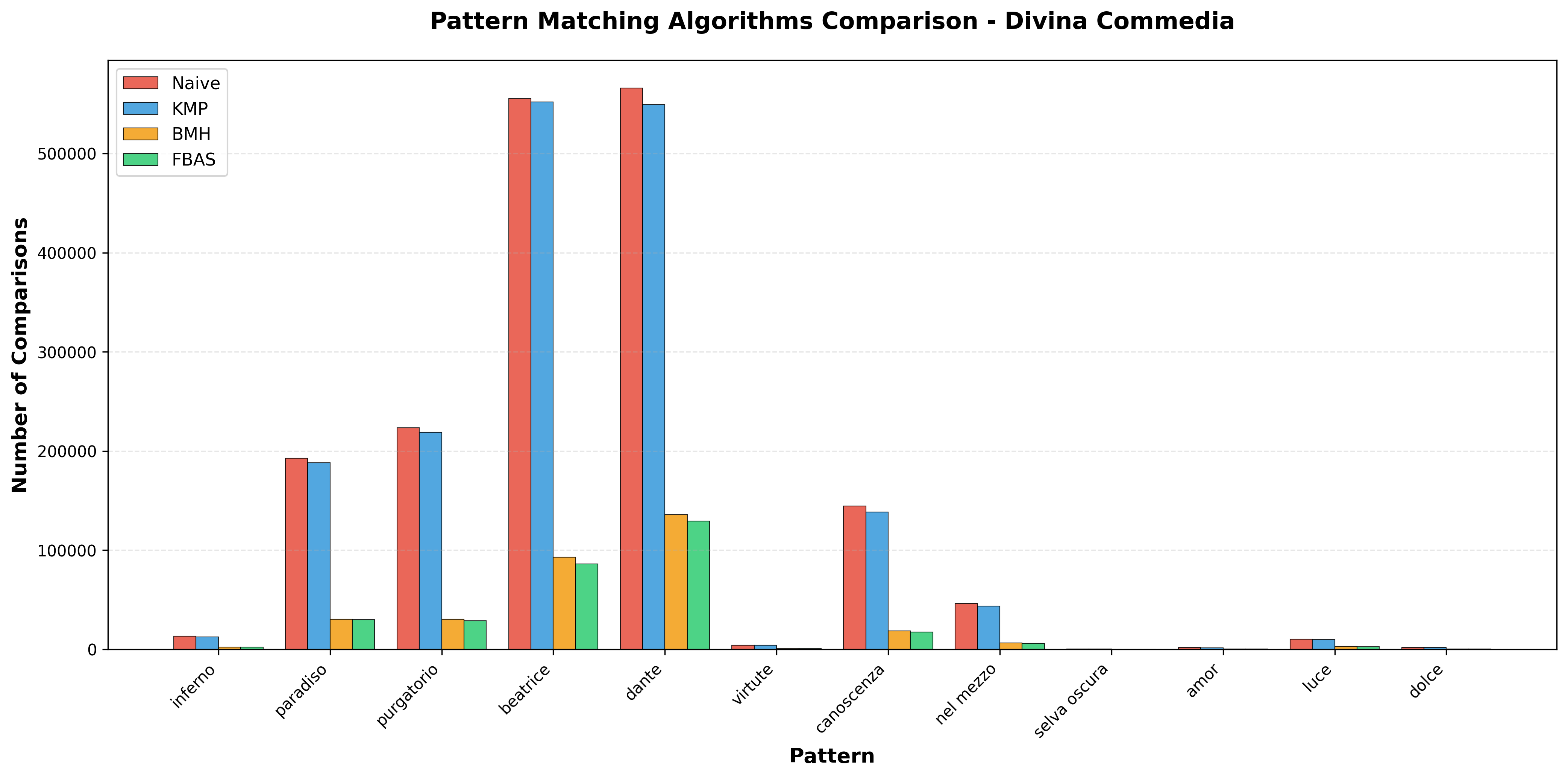}}
\caption{Comparison of character comparisons across all four algorithms. Note the logarithmic scale highlighting the efficiency of shift-based approaches.}
\label{fig:comparison_all}
\end{figure}

Figure~\ref{fig:comparison_all} illustrates the dramatic performance advantage of both BMH and FBAS over naive and KMP approaches. The logarithmic scale reveals that shift-based algorithms reduce comparisons by an order of magnitude.

\begin{figure}[htbp]
\centerline{\includegraphics[width=\columnwidth]{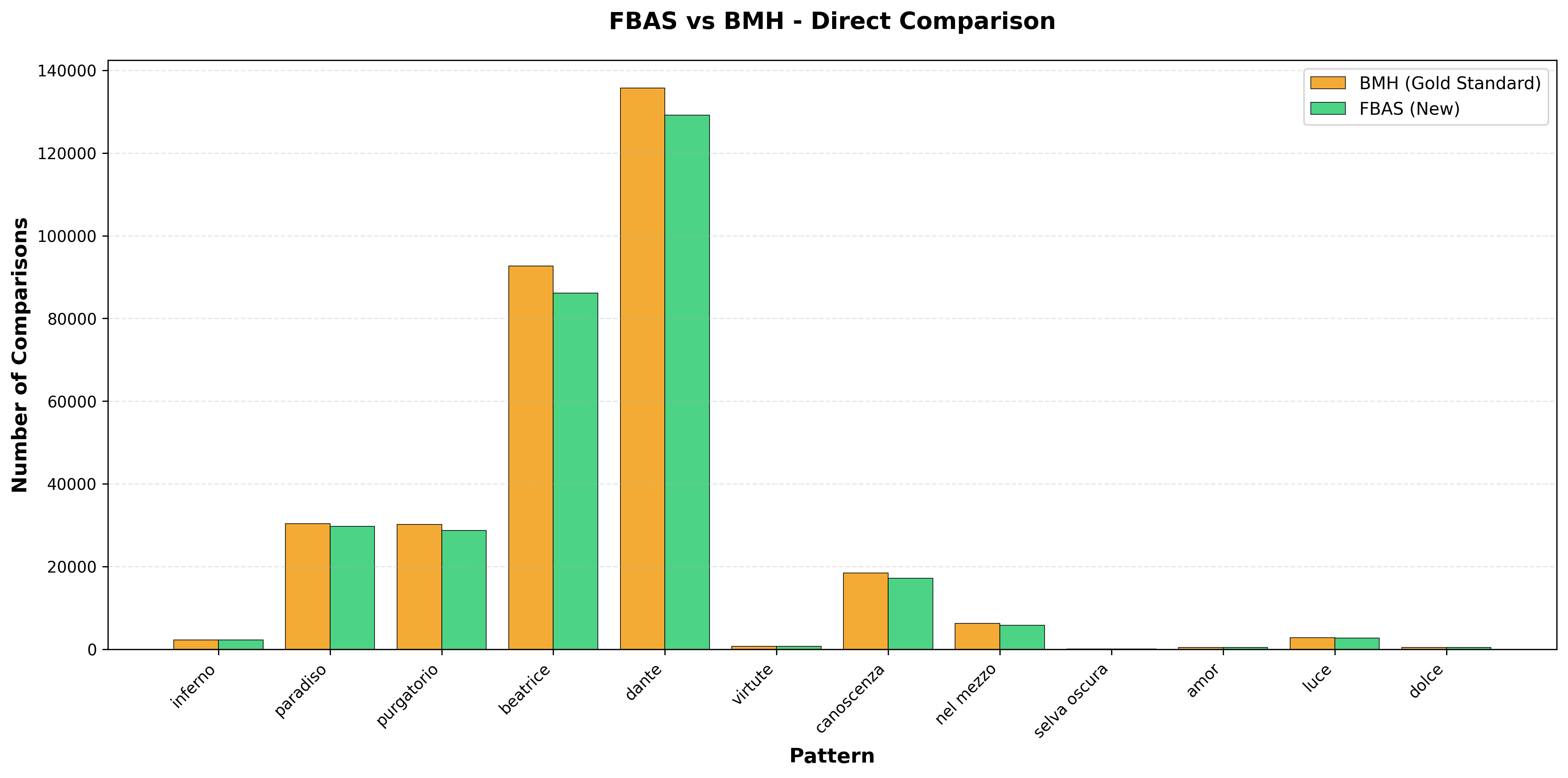}}
\caption{Direct comparison between FBAS and BMH algorithms. FBAS consistently requires fewer comparisons across all tested patterns.}
\label{fig:fbas_vs_bmh}
\end{figure}

Figure~\ref{fig:fbas_vs_bmh} provides a detailed view of the FBAS versus BMH comparison. While both algorithms perform similarly on patterns with only common characters, such as ``inferno'', FBAS shows significant advantages on patterns containing rare characters like 'z', 'b', and 'g'.

\subsection{Analysis}

\subsubsection{Impact of Character Rarity}
\begin{figure}[htbp]
\centerline{\includegraphics[width=\columnwidth]{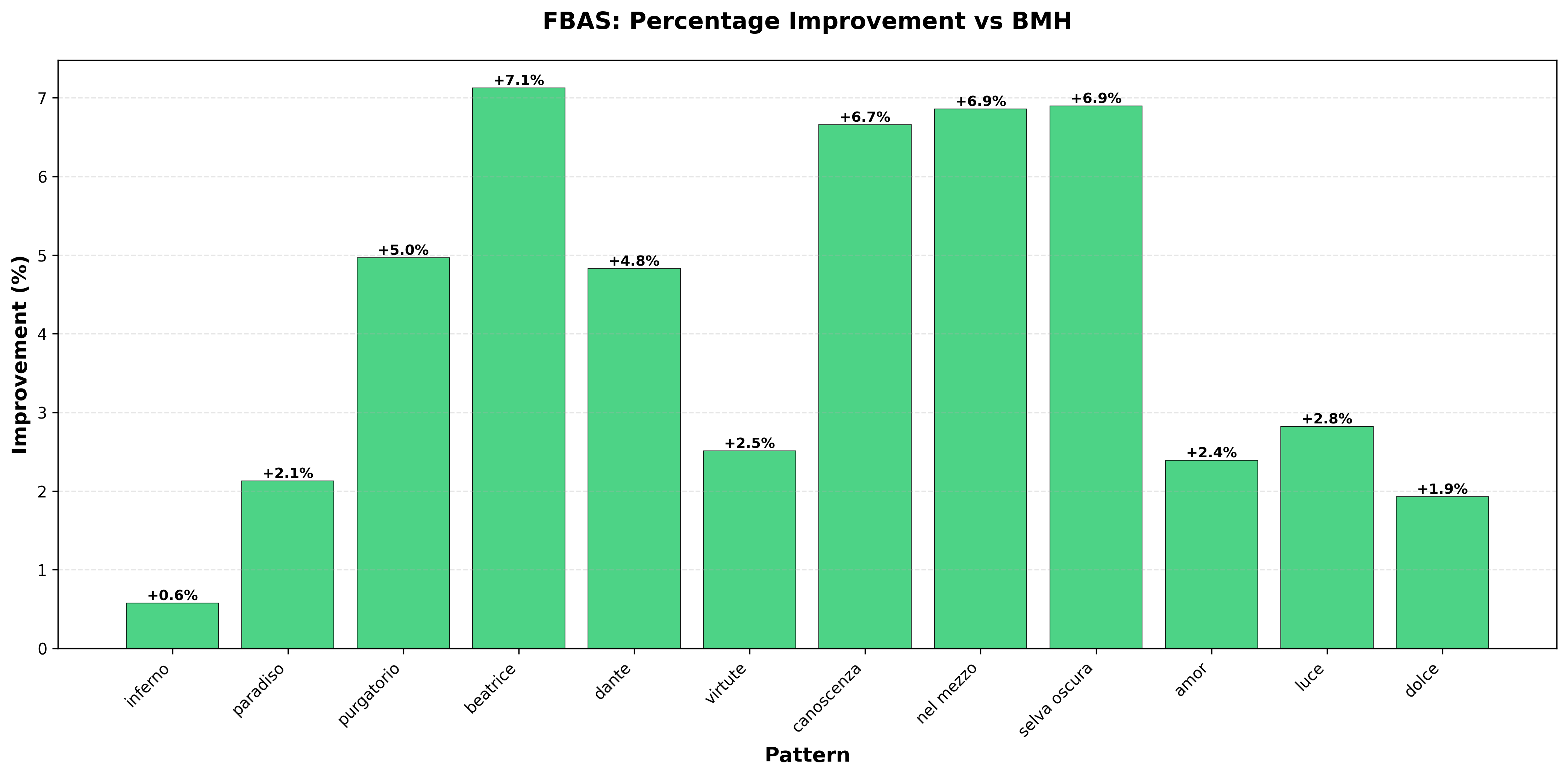}}
\caption{Percentage improvement of FBAS over BMH for each pattern. Patterns with rare characters (z, b, g) show the highest gains.}
\label{fig:improvement}
\end{figure}

Figure~\ref{fig:improvement} reveals a clear correlation between improvement percentage and the presence of rare characters in the pattern. Patterns containing 'z', the rarest character in our frequency table, consistently show improvements above 6\%. The pattern ``beatrice'' achieves 7.12\% improvement with anchor 'b' (rarity score 10), ``selva oscura'' reaches 6.90\% with anchor 'u' (rarity score 16), and ``nel mezzo'' shows 6.86\% with anchor 'z' (rarity score 1). This validates our hypothesis that selecting statistically improbable characters as anchors accelerates the fail-fast mechanism.

\subsubsection{Aggregate Performance}
Across all 12 patterns, FBAS performs 303,194 total comparisons compared to BMH's 320,275, representing a 5.33\% overall reduction. More significantly, FBAS wins in 12 out of 12 patterns, demonstrating consistent superiority regardless of pattern characteristics.

\begin{figure}[htbp]
\centerline{\includegraphics[width=\columnwidth]{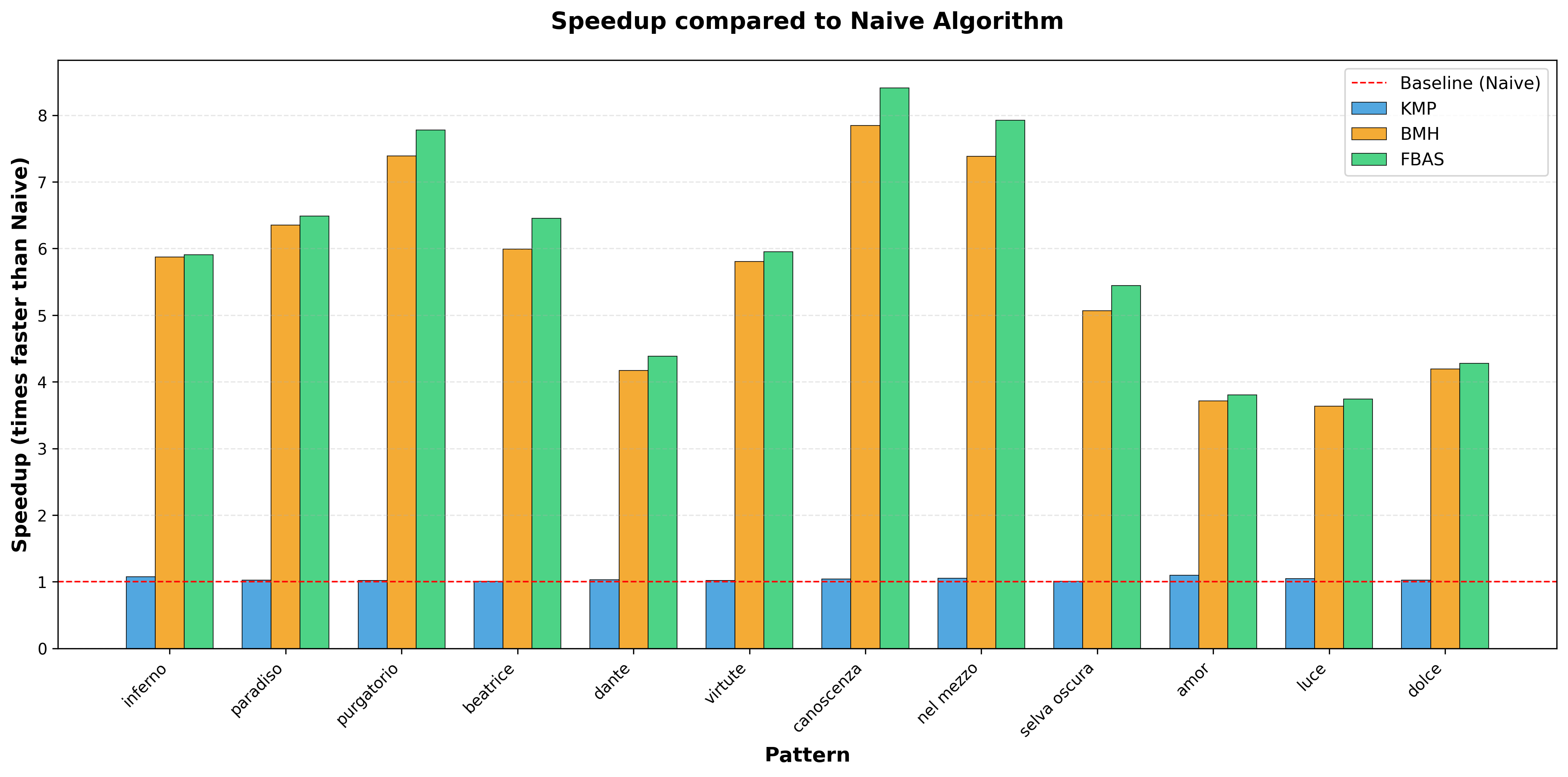}}
\caption{Speedup factor relative to naive search. Both BMH and FBAS demonstrate order-of-magnitude improvements, with FBAS maintaining a consistent edge.}
\label{fig:speedup}
\end{figure}

Compared to naive search, both FBAS and BMH show dramatic improvements. Figure~\ref{fig:speedup} illustrates the speedup factor relative to naive search, where FBAS achieves an average 5.8× speedup, or 81.70\% fewer comparisons.

\subsubsection{Pattern Length Effects}
Longer patterns generally benefit more from FBAS optimization. The pattern ``purgatorio'' (length 10) shows 4.97\% improvement, while shorter patterns like ``inferno'' (length 7) show only 0.58\%. This aligns with our theoretical analysis: longer patterns provide more opportunities to identify rare anchor characters and achieve larger shifts.

\subsubsection{Practical Implications}
The consistent 2-7\% reduction in character comparisons translates directly to performance gains in real-world applications. For text search operations processing large corpora or performing frequent searches, these savings accumulate significantly. The algorithm's simplicity, having identical shift heuristics to BMH with only a modified verification order, means it can be easily integrated into existing systems without architectural changes.

\subsection{Limitations and Future Work}

While our results demonstrate clear advantages for Italian text, we acknowledge several areas for future investigation:

\begin{itemize}
    \item {Language dependency:} Our frequency table is optimized for Romance languages. Testing on Germanic, Slavic, or logographic writing systems would reveal language-specific effects.
    \item {Dynamic frequency adaptation:} Computing character frequencies from the actual text corpus rather than using predefined tables could further optimize anchor selection.
    \item{Multiple occurrences:} Our metric counts all comparisons including those after the first match. Analyzing time-to-first-match separately would provide additional insight.
    \item{Cache effects:} Modern CPU architectures exhibit complex cache behavior. Microbenchmarking on various hardware platforms would complement our comparison-based analysis.
\end{itemize}

\section{Conclusion}

We have presented FBAS, a frequency-based enhancement to the Boyer-Moore-Horspool algorithm that leverages statistical character rarity to optimize pattern verification. By preprocessing the pattern to identify its least frequent character as an anchor and prioritizing its verification during search, FBAS achieves a fail-fast mechanism that reduces unnecessary comparisons without modifying the shift heuristics or increasing algorithmic complexity.

Our experimental evaluation on Dante's \textit{Divina Commedia} demonstrates consistent improvements across all 12 tested patterns, with FBAS requiring 5.33\% fewer character comparisons than standard BMH (303,194 vs. 320,275 total comparisons). The algorithm achieves 100\% win rate (12/12 patterns) and shows particularly strong performance on patterns containing rare characters, with improvements reaching 7.12\% for ``beatrice'' and 6.90\% for ``selva oscura''. Compared to naive search, FBAS reduces comparisons by 81.70\%, achieving an average 5.8× speedup.

The key strength of FBAS lies in its simplicity and practicality. It maintains the $O(m)$ preprocessing and $O(n/m)$ average-case complexity of BMH while requiring no additional asymptotic space overhead. The modification is minimal, consisting of a single preprocessing step to identify the anchor and a reordered verification loop, making it straightforward to integrate into existing codebases. For applications performing frequent searches over natural language corpora, these consistent 2-7\% efficiency gains translate to meaningful performance improvements with negligible implementation cost.

Future work includes evaluating FBAS on diverse language families beyond Romance languages, investigating dynamic frequency adaptation based on actual text statistics, and exploring multi-anchor strategies that verify multiple rare characters in parallel. The success of this approach suggests that incorporating domain-specific statistical knowledge into classical algorithms remains a fruitful avenue for practical optimization.

\end{document}